\documentclass[lettersize,journal]{IEEEtran}
\IEEEoverridecommandlockouts
\usepackage{cite}
\usepackage{amsmath,amssymb,amsfonts}
\usepackage{algorithmic}
\usepackage{graphicx}
\usepackage{textcomp}
\usepackage{xcolor}
\usepackage{balance}
\newtheorem{theorem}{Theorem}

\newtheorem{remark}[theorem]{Remark}
\newtheorem{definition}[theorem]{Definition}

\def\BibTeX{{\rm B\kern-.05em{\sc i\kern-.025em b}\kern-.08em
    T\kern-.1667em\lower.7ex\hbox{E}\kern-.125emX}}

\begin{document}

\title{On The Capacity of Low-Rank Dyadic Fading Channels in the Low-SNR Regime
\thanks{This research work is supported by the Science and Engineering Research Board (SERB), Government of India grant: TAR/2022/000194.\vspace*{-1em}}
}



\author{
\IEEEauthorblockN{Kamal Singh\IEEEauthorrefmark{1}}
\thanks{

\IEEEauthorrefmark{1}{The author is with the Department of Electrical Engineering, Shiv Nadar Institution of Eminence, Delhi NCR, India (e-mail: kamal.singh@snu.edu.in).}}
\vspace*{-0.75cm}
}

%

\maketitle
\begin{abstract}
We characterize the capacity of a low-rank wireless channel with varying fading severity at low signal-to-noise ratios (SNRs). The channel rank deficiency is achieved by incorporating pinhole condition. The capacity degradation with fading severity at high SNRs is well known: the probability of deep fades increases significantly with higher fading severity resulting in poor performance. Our analysis of the dyadic pinhole channel at low-SNR shows a very counter-intuitive result that - \emph{higher fading severity enables higher capacity at sufficiently low SNR}. The underlying reason is that at low SNRs, ergodic capacity depends crucially on the probability distribution of channel peaks (tail distribution); for the pinhole channel, the tail distribution improves with fading severity. This allows a transmitter operating at low SNR to exploit channel peaks `more efficiently' and hence improves spectral efficiency. We derive a new key result quantifying the above dependence for the double-Nakagami-$m$ fading pinhole channel - the capacity ${C} \propto (m_T m_R)^{-1}$ at low SNR, where $m_T m_R$ is the severity parameters (product) of the fadings involved.
\end{abstract}
\begin{IEEEkeywords}
Double-fading, pinhole, tail distribution, low-rank channel, low SNR.
\vspace*{-0.25cm}
\end{IEEEkeywords}
\section{Introduction}

In this work, we consider low-power short-range wireless communications through a low-rank fading channel - a bonafide use case in many communication scenarios requiring simple wireless connectivity with much relaxed constraints on throughput and data latency. This is certainly true, for instance, in low-complexity wireless channels in the low-rate wireless personal area networks (LR-WPANs). Low-rate communication on control channels in wireless networks is another relevant example. More example applications may include low-rate wireless links in low-power IoT networks, low-power sensor networks, etc. 

For the important class of~\emph{cascaded or multi-hop} channels where radio wave propagation occurs in succession through multiple clusters (layers) of scatterers local to each node, the overall (end-to-end) fading is modeled as the `product' of the independent per-cluster or per-hop fadings. Cascaded channels have attracted considerable research attention in the past decade due to their wide range of applications in emerging direct device-to-device wireless communication systems. Select relevant prior works on product fading channels include degenerate pinhole channels in MIMO systems\!\cite{chizhik2}\!\cite{shinlee}, relay communications\!\cite{dohler}, backscatter RF channels\!\cite{arnitz}\!\cite{sroy}, double-Rayleigh fading channel for the urban area when both transmitter and receiver show large-scale mobility\!\cite{gesbert}\!\cite{fortune}\!\cite{kovacs}, and, more recently, reconfigurable intelligent surface (RIS) assisted dual-hop wireless channels where the RIS elements are modeled as keyholes/scatterers distributed between the transmitter and receiver\!\cite{arslan}\!\cite{yang}, etc. Interested readers can further explore\!\cite{dohler}\!\cite{Nnakagami}\!\cite{nidhi} for more details on the statistical characterization and modeling aspects of some popular product fading channels.

\begin{figure}[htbp]
\centering
\includegraphics[scale = 1.2]{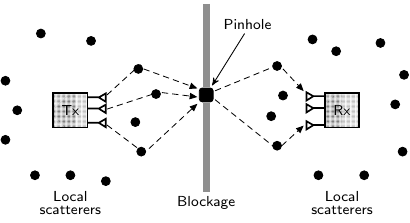}
\vspace*{-0.13em}
\caption{Illustration of signal propagation through two local rich-scattering wireless environments connected via a pinhole.}\label{fig:pinhole}
\vspace*{-0.5em}
\end{figure}
For now, we consider a single-antenna Tx/Rx system with dyadic or dual-hop forward scatter signal propagation that occurs via a pinhole as shown in Fig.~\ref{fig:pinhole}, modeling a large-scale obstruction between the source and destination; hence forward-wave propagation is possible only via passing through a waveguide or `pinhole'. Theoretical bases for the degenerate pinhole effect in wireless channels are proposed in\!\cite{chizhik2}\!\cite{gesbert}, whereas empirical validations of the pinhole in controlled indoor environments are confirmed in\!\cite{almers0}\!\cite{almers}.
The captured incident energy at the pinhole is re-radiated towards the destination and represents a rank-1 end-to-end channel that is doubly fading (or dyadic), described as 
\begin{align}\label{eq:pinholeH22}
{h} := {h}_1 {h}_2
\end{align}
where ${h}_1$ and ${h}_2$ denote the \emph{complex} scalar channel coefficients from the source-to-pinhole and pinhole-to-destination respectively. 
The product fading implies \emph{worsened} channel statistics\!\cite{gesbert}; for example, the probability that the channel is in \emph{deep fade}\footnote{Precisely, a `deep fade' event occurs when the channel gain (i.e., squared fading envelope) is low; e.g., values much less than the mean channel gain.} (i.e., the overall channel gain is less than a certain threshold, say, $\gamma_{\text{th}}$) increases, i.e.,
\begin{align}\label{eq:pinholeH33}
\,\,\,\,\,\,\,\,\,\text{Prob}(|{h}|^2 \leq  \gamma_{\text{th}}) \, >  \, \text{Prob}(|h_i|^2  \leq  \gamma_{\text{th}}), i =1,2
\end{align}
suggesting that the fading becomes more severe for the  cascade as compared to single-hop. 

\begin{figure}[htbp]
\centering
\includegraphics[width=0.495\textwidth]{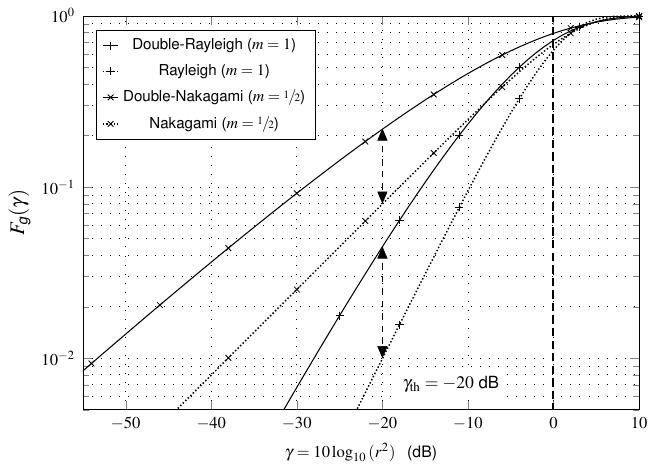}
\caption{Cumulative distribution function $F_{g}(\cdot)$ of single \& dyadic Nakagami-$m$ channel gains. The channel gain $g := r^2$ is the squared fading envelope, i.e., $r = |h|$ for dyadic and $r = |h_i|$ for single-hop channels (see~\eqref{eq:pinholeH22}). The mean channel gain is normalized to unity (i.e., $\mathbb{E} \,[g] = 1$ or $0$ dB) in all cases.}\label{fig:CDF}
\vspace*{-1.35em}
\end{figure}
\noindent
The probabilities of deep fade for a few cases of single-hop and dyadic Nakagami-$m$ fading channels are illustrated in Fig.~\ref{fig:CDF}. For $m=1$ (Rayleigh), the probability of fade depth exceeding $20$ dB (relative to the $0$ dB mean) has increased from $1$\% to $4$\% for single-hop to double-hop. In contrast for Nakagami fading with $m = 0.5$, the probability for the same $20$ dB or more fade increases from $8$\% to $20$\% as the cascade level is changed from single to dyadic. In short, the dyadic Nakagami-$m$ model allows us to parametrically explore the impact of \emph{change in fading severity level}.

Mobile wireless channels typically operate under average and peak power constraints, finite battery capacity at the mobile terminals, etc. - and thus achieving an adequate signal-to-noise ratio (SNR) at the receiver under dynamic channel conditions can often be very challenging. Specifically, the spectral efficiency is significantly limited in power-limited operations. Improving spectral efficiency of RF wireless data transmission in the low power regime and exploration of consequent spectral efficiency limits is a direction of continuing interest; see\!\cite{lozano}\!\cite{tall}\! \cite{rezki}. The existing literature on \emph{single-hop} wireless fading channels and their performance limits at  low SNRs provides a couple of interesting observations as follows:
\begin{itemize}
\item In the low-SNR regime, the key to improving the fading channel capacity lies in the efficient adaptation of transmit power based on (assumption of) perfect channel knowledge at the source\!\cite{tsebook}.
\item Low-SNR (ergodic) capacity results of some scalar fading channels latently allude to the possibility that a higher fading severity level provides opportunities to exploit for capacity gains at low SNRs\!\cite{rezki}. 
\end{itemize}
\noindent
These observations together lead to an interesting outstanding question: whether it is possible for the dyadic pinhole channel in the power-limited regime to exploit the increased channel fading severity on offer. In this paper, we will show that a wide class of dyadic fading channels (relay/keyhole, etc.) can leverage higher fading severity in dyadic (by extension, multi-hop) transmissions for capacity enhancement at low SNRs.

The following notation is used throughout: $\mathbb{E}_{{x}} [\cdot]$ denotes expectation and $f_{{x}} (\cdot)$ the corresponding probability density function (PDF) with respect to the random variable (RV) ${x}$. The $\log(\cdot)$ function represents the natural logarithm and $\mathrm{max}\{0,z\}$ is denoted by $z^{+}$.

\section{Dyadic/Double-fading (Pinhole) Channel }\label{sec:two}

We restrict our attention to the double-fading channel in Fig.~\ref{fig:pinhole} with perfect channel side information (CSI) at both sides and \emph{frequency-flat} independent Nakagami-$m$ fadings for the two locally rich scattering environments connected via the pinhole. The primary reason for selecting the Nakagami-$m$ distribution is its flexibility in capturing varying severity levels via its free fading parameter $m$ and a consequent good fit to empirical measurements\!\cite{Nakagami_dist}. 
The transmitted signal can propagate to the receiver side \emph{only} via the pinhole connecting the two rich scattering environments.

\vspace*{0.15em}
With $B$ as the \emph{channel bandwidth} and $S$ as the \emph{long-term average power constraint}, we represent the discrete-time complex baseband symbol at the receiver as follows (the time index is suppressed for simplicity):
\begin{align}\label{eq:sys_model}
{y} = {h} {x} + {w}
\end{align}
where ${h}$ is the double-fading channel coefficient, ${x}$ is the transmitted symbol, and ${w} \sim \mathcal{CN} (0, N_0)$ is additive complex Gaussian noise. Notice that the transmit power constraint per complex symbol time is $S/B$  while the noise variance per complex symbol time is $N_0$. For simplicity, we assume an equivalent channel model with the noise variance $N_0$ set to unity and the transmit power constraint as $S/(N_0 B)$.
\begin{definition}\label{eq:def:SNR}
The average signal-to-noise ratio is defined as
\begin{align}
\mathrm{SNR} := S/(N_0 B).
\end{align}
\end{definition}

For a pinhole that acts as an ideal scatterer, the composite double-fading channel is\footnote{We consider short-range wireless communications; thus, the impact of large-scale channel effects is ignored here.}
\begin{align}\label{eq:pinholeH}
{h} := {h}_R \, {h}_T
\end{align}
where ${h}_T := {\alpha} e^{j {\psi}}$ and ${h}_R := {\beta} e^{j {\phi}}$ denote the channel coefficients for the source-to-pinhole and pinhole-to-destination links respectively. We assume that the phase $\psi$ is uniformly distributed in $[0,2\pi)$ and the fading envelope $\alpha$ admits a Nakagami-$m$ PDF as follows:
\begin{align}\label{eq:dist:naka1}
f_{\alpha} (\alpha) &= \dfrac{2}{\Gamma(m_T)}\left(\dfrac{m_T}{\Omega_T}\right)^{m_T} \alpha^{2m_T - 1} e^{-\tfrac{m_T}{\Omega_T}\alpha^2},\,\,\alpha \geq 0
\end{align}
where $m_T \geq {1}/{2}$ and $\Omega_T > 0$ are the fading and scale parameters of the Nakagami-$m$ distribution respectively\!\cite{Nakagami_dist}, and $\Gamma(\cdot)$ is the Gamma function\!\cite{tableofintegrals}. Likewise, an independent
Nakagami-$m$ PDF with $(m_R,\Omega_R)$ parameters characterizes the fading envelope $\beta$ of the channel coefficient ${h}_R$ as follows:
\begin{align}\label{eq:dist:naka2}
f_{{\beta}} (\beta) &= \dfrac{2}{\Gamma(m_R)}\left(\dfrac{m_R}{\Omega_R}\right)^{m_R} \beta^{2m_R - 1} e^{-\tfrac{m_R}{\Omega_R}\beta^2},\,\,\beta \geq 0.
\end{align}
Note that $\Omega_T = \mathbb{E}_{{\alpha}} [\alpha^2]$ and $\Omega_R = \mathbb{E}_{{\beta}} [\beta^2]$ are not necessarily identical. The fading parameters $m_T$ and $m_R$ signify the severity of the envelope attenuation on the two links (or hops) with $m_R = 1 = m_T $ representing the double-Rayleigh baseline [values smaller or larger than one indicate fading conditions more or less severe than double-Rayleigh fading, see Fig.~\ref{fig:CDF}].

\subsection{Full CSI Dyadic (Pinhole) Channel Capacity at Low SNR}\label{sec:three}
The ergodic capacity of the double-fading channel ${h}$ with `perfect channel state information at both ends' (full CSI) can be computed as follows\!\cite{tsebook}:
\begin{align}
C &= \mathbb{E}_{{h}} \,[\log(1 + {h} P({h}){h}^{*})] 
= \mathbb{E}_{{\lambda}} \,[\log(1 +{\lambda} P({\lambda}))]\label{eq:cap:csit_do}
\end{align}
where ${\lambda}:= |{h}_T|^2  |{h}_R|^2$ is the overall channel gain and $P({\lambda})$ is the transmit power control satisfying the constraint:
\begin{align}\label{eq:power_const_a}
\mathbb{E}_{{\lambda}} \,[P({\lambda})] \,=\, \mathrm{SNR}.
\end{align}

Since $|{h}_T|^2 = \alpha^2$ and $|{h}_R|^2 = \beta^2$, the independent squared Nakagami-$m$ variates $\alpha^2$ and $\beta^2$ are Gamma distributed,\footnote{A non-negative RV denoted by $z \sim  \Upsilon(\Omega,m)$ is Gamma distributed with its PDF described as $f_z(z) = z^{m-1} e^{-z/\Omega} /({\Gamma(m) \Omega^m})$, where $m > 0$ and $\Omega > 0$ are the shape and scale parameters respectively; see~\cite{Zwillinger} for details.} i.e., $\alpha^2 \sim \Upsilon(\Omega_T/m_T,m_T)$ and $\beta^2 \sim \Upsilon(\Omega_R/m_R,m_R)$. Thus, the PDF of the channel gain ${\lambda}$ (for $\lambda > 0$) is computed as follows:
\begin{align}
f_{{\lambda}} (\lambda) &= \int_{0}^{\infty} f_{|{h}_T|^2} (z) \, f_{|{h}_R|^2} \left({\lambda}/{z}\right)\,\, d\,\mathrm{ln}(z)\notag\\
&= A \cdot K_{m_R - m_T} \left(2\sqrt{\dfrac{\lambda}{b_{TR}}}\,\right)
\left(\dfrac{\lambda}{b_{TR}}\right)^{\tfrac{m_T + m_R}{2} - 1}\label{eq:cap:csit0}
\end{align}
where, for convenience of notation, we have introduced $b_{TR} := {\Omega_T \Omega_R}/{(m_T m_R)}$, $A := 2/(b_{TR} \Gamma(m_R) \Gamma(m_T))$, and $K_{\nu}(\cdot)$ is the $\nu$-th order Bessel function of the second kind~\cite{tableofintegrals}.

\vspace*{0.2em}
For the scalar fading channel ${\lambda}$, the classical water-filling formula given by $ P(\lambda) = \left({1}/{\lambda_0} - {1}/{\lambda}\right)^+$ is the optimal power solution~\cite{tsebook}. Thus, positive power is allocated only when the fade level is above the channel cutoff $\lambda_0$. The channel cutoff $\lambda_0$ is chosen such that the transmit power constraint~\eqref{eq:power_const_a} is satisfied. Substituting $P(\lambda)$ in~\eqref{eq:cap:csit_do}, we get
\begin{align}
C &= \int_{\lambda_0}^{\infty} \log \left(\dfrac{\lambda}{\lambda_0}\right) f_{{\lambda}}(\lambda)  d \lambda\label{eq:cap_der_do}.
\end{align}
The power constraint~\eqref{eq:power_const_a} becomes:
\begin{align}\label{eq:cutoff1}
\mathrm{SNR} = \int_{\lambda_0}^{\infty}\left(\dfrac{1}{\lambda_0} -\dfrac{1}{\lambda}\right) f_{\lambda}(\lambda) d \lambda
\end{align}
From~\eqref{eq:cutoff1}, it is straightforward to see that the channel cutoff $\lambda_0$ varies inversely with the $\mathrm{SNR}$. In particular, with some effort (using~\eqref{eq:cap:csit0}), it can be shown that $\lambda_0 \to \infty$ as $\mathrm{SNR} \to 0$.

\vspace*{0.2em}
Unfortunately, the presence of the complicated squared double-Nakagami-$m$ distribution function $f_{\lambda}(\lambda)$ (see~\eqref{eq:cap:csit0}) poses analytical difficulties in solving the integral~\eqref{eq:cap_der_do} for an exact closed form for the channel capacity $C$. As an alternate approach, {\em asymptotic analysis in the extreme SNR regimes} can provide valuable insights as it reveals dependence that is, in general, broadly applicable for moderate SNR conditions. 

\vspace*{0.2em}
Specifically, we will analyze the capacity behavior of the double-fading channel in the asymptotic limit of low SNR (i.e., $\mathrm{SNR} \to 0$). For this purpose, we consider defining any two functions, say $f(u)$ and $g(u)$, to be asymptotically equivalent (in the limit as $u  \to  0 $) as follows:\vspace*{0.35em}
\begin{definition}\label{eq:def:approx}
$f \approx g\,\,$ if and only if $\,\,\,\lim\limits_{u \to 0} \,\,\, \frac{f(u)}{g(u)} \,=\, 1$.\vspace*{0.1em}
\end{definition}
\vspace*{0.5em}
\noindent
The following theorem states our main result.\vspace*{0.35em}
\begin{theorem}\label{eq:theorem3}
Under full CSI, the asymptotic (low-SNR) capacity of the double-fading pinhole channel subjected to independent Nakagami-$m$ fadings between the source-to-pinhole and pinhole-to-destination side respectively [described by~\eqref{eq:sys_model}$/$\eqref{eq:pinholeH}] is given by
\begin{align}\label{eq:thm_upper_part}
C \approx \left(\dfrac{\Omega_T \Omega_R}{ m_T  m_R}\right) \dfrac{{\mathrm{SNR}}}{4}  \log^2 \left(\dfrac{1}{\mathrm{\mathrm{SNR}}}\right).
\end{align}
\end{theorem}
\vspace*{0.2em}
\begin{IEEEproof}
Recall that as $\mathrm{SNR} \to 0$, $\lambda_0 \to \infty$. Thus, at low SNRs, the source's transmitter allocates power \emph{only} for very high channel gains. Hence, 
the first step to simplify~\eqref{eq:cap_der_do} at low SNR is to approximate the distribution function $f_{{\lambda}}(\lambda)$ (see~\eqref{eq:cap:csit0}) for very high channel gains. To do this, it is convenient to consider the series representation (for large argument) of the modified Bessel function of the second kind given below~\cite{tableofintegrals}:
\begin{align}\label{eq:bessel_approx}
\phantom{x}K_v (t)  \, \approx \, \sqrt{\frac{\pi}{2t}} \,e^{-t} + o\left(\frac{1}{t}\right).
\end{align}
Substituting~\eqref{eq:bessel_approx} in~\eqref{eq:cap_der_do} yields
\begin{align}
C \,&\approx\, K \, \cdot \, \int_{\mu_0}^{\infty} \log \left(\dfrac{{\lambda}}{\mu_0}\right) \,\lambda^{\frac{m_T + m_R}{2} - \frac{5}{4}} \, e^{-2\sqrt{\lambda}}\,\, d \lambda\label{eq:cap_der_do222}
\end{align}
where $\mu_0 := {\lambda_0}/{b_{TR}}$ is the `scaled' channel cutoff and $K := \sqrt{\pi}/({\Gamma(m_T)\Gamma(m_R)})$. To proceed further to approximate the integral in~\eqref{eq:cap_der_do222}, we borrow the identity given below~\cite{tableofintegrals}:
\begin{align}
\int_{\eta}^{\infty} \log \left(\frac{t}{\eta}\right) t^{a}  e^{-2\sqrt{t}} \, d t= \dfrac{1}{4^{a}}{G_{2,3}^{3,0}\left({2\sqrt{\eta}}
\middle\vert\hspace{-0.5em}
\begin{array}{c}
1,1\\
0,0,2(a + 1)\\
\end{array}\hspace{-0.5em}\right)}\label{eq:cap_der_do3}
\end{align}
where $G(\cdot)$ is the  Meijer's $G$ function~\cite{tableofintegrals}. 
The series representation of the Meijer's $G$ function for large 
argument is expressed as follows~\cite{tableofintegrals}:
\begin{align}\label{eq:cap_der_do323}
\phantom{c}G_{2,3}^{3,0}&\left({2\sqrt{\eta}}
\middle\vert\hspace{-0.5em}
\begin{array}{c}
1,1\\
0,0,2(a + 1)\\
\end{array}\hspace{-0.5em}\right) \,\,\approx \,\, 4^{a} \,{e^{-2\sqrt{\eta} }}\,\, {\eta^{a - 1}}\notag\\
\phantom{xxx}\times\,&\biggl[\,\eta \,+\,  \frac{(1 \,+\,12 a^2 \,+\, 8a\sqrt{\eta})}{4} \,+\,  \frac{1}{4^a} o\left(\frac{1}{\eta}\right)^{\frac{3}{2}}\,\biggr].
\end{align}
Simplifying~\eqref{eq:cap_der_do3} by taking only the first (largest) term in~\eqref{eq:cap_der_do323}, then~\eqref{eq:cap_der_do222} is approximated as
\begin{align}
C \, \approx \, K \, \cdot \, {\mu_0}^{\frac{m_T + m_R}{2} - \frac{5}{4}} \, e^{-2\sqrt{\mu_0}}\,.\label{eq:cap_der_do4}
\end{align}

To derive $C-\mathrm{SNR}$ relation explicitly, we need to extract the $\mu_0-\mathrm{SNR}$ dependence at low SNRs from the power constraint~\eqref{eq:cutoff1}. We repeat the $f_{{\lambda}}(\lambda)$ low-SNR approximation (using~\eqref{eq:bessel_approx}) in~\eqref{eq:cutoff1} to produce
\begin{align}\label{eq:cutoff121}
\mathrm{SNR} \left(b_{TR}\right) \,&\approx \, K \,\cdot \, \left[\, \dfrac{I_1 (\mu_0)}{\mu_0}  - I_2 (\mu_0) \, \right]
\end{align}
where
\begin{align}\label{eq:I_definitions}
\begin{cases}
I_1 (\mu_0) \,=\,\,\,\,\Gamma\bigl(b + 1, 2\sqrt{\mu_0}\bigr)/2^{b}\\[0.35em]
I_2 (\mu_0) \,=\, 4\Gamma\bigl(b - 1, 2\sqrt{\mu_0}\bigr)/2^{b}
\end{cases}
\end{align}
\noindent
where, in turn, $b := m_T + m_R - \frac{3}{2}$, and the $\Gamma(\cdot,\cdot)$ is the upper incomplete Gamma function whose power series expansion at large input is given below~\cite{tableofintegrals}:
\begin{align}\label{eq:gamma_approx}
\Gamma(s,t) \, \approx \,t^s e^{-t}  \biggl(\dfrac{1}{t} + \dfrac{s-1}{t^2} + o\left(\dfrac{1}{t}\right)^3\biggr).
\end{align}
By considering only the first two largest terms in~\eqref{eq:gamma_approx},~\eqref{eq:cutoff121} gets simplified to
\begin{align}\label{eq:cutoff131}
\mathrm{SNR} \left(b_{TR}\right) \,&\approx\, K \, \cdot \, {\mu_0}^{\frac{m_T + m_R}{2} - \frac{9}{4}} \, e^{-2\sqrt{\mu_0}}.
\end{align}
Comparing~\eqref{eq:cap_der_do4} and~\eqref{eq:cutoff131}, we can easily deduce that
\begin{align}\label{eq:Cap_lamda}
C \, \approx \, \mu_0 \left(\frac{\Omega_T \Omega_R}{ m_T  m_R}\right)\, \mathrm{SNR}.
\end{align}
\noindent
To complete the log characterization of the capacity $C$ as given in~\eqref{eq:thm_upper_part}, we solve~\eqref{eq:cutoff131} for $\mu_0$ (the scaled channel cutoff) by applying the natural logarithm to both sides and ignoring the smaller terms to give
\begin{align}\label{eq:n_below1223}
\log (\mathrm{SNR})  \approx -2 \sqrt{\mu_0} \,\,\, \Rightarrow \,\,\mu_0   \,\approx\, \dfrac{1}{4}\log^2\Bigl(\dfrac{1}{\mathrm{SNR}}\Bigr).
\end{align}
Substituting~\eqref{eq:n_below1223} back into~\eqref{eq:Cap_lamda} completes the proof.
\end{IEEEproof}

The essence of Theorem~\ref{eq:theorem3} is that it provides a simple and clear characterization of the impact of fading severity on the dyadic channel capacity at low-SNRs - that ${C} \propto (m_T m_R)^{-1}$ where $m_T m_R$ is the product of fading severity parameters of the two independent (per-hop) Nakagami-$m$ fadings involved.
\begin{remark} Extending the above asymptotic analysis to a \emph{multi-antenna pinhole channel} with $n_T$ transmit and $n_R$
receive antennas with i.i.d. Nakagami-$m$ fadings for all the signal paths is straightforward. The low-SNR capacity result in~\eqref{eq:theorem3} holds with the only modification that the overall mean path gain ($\Omega_T \Omega_R$ earlier) is improved by the factor $n_R n_T$. 
\end{remark}
\section{Numerical Results \& Discussion}
\begin{figure}[htbp]
\centering
\vspace*{-1.em}
\includegraphics[width=0.5\textwidth]{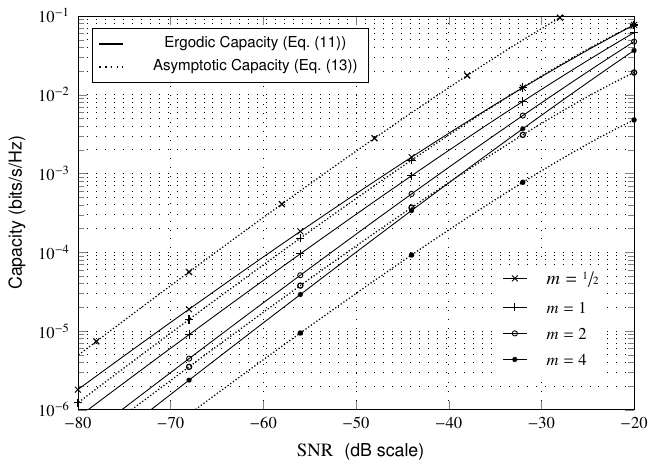}
\vspace*{-1.75em}
\caption{Comparison of exact \& asymptotic capacity of the double-Nakagami-$m$ pinhole channel at low SNRs. For simplicity, $m_R = m_T$ (say $m$) and $\Omega_R = \Omega_T = 1$.}\label{fig:one}
\end{figure}

The validity of the asymptotic approach to analyze the low-SNR capacity of the dyadic pinhole channel (i.e., Theorem~\ref{eq:theorem3}) is verified in Fig.~\ref{fig:one}. The exact capacity approaches the asymptotic capacity in shape at sufficiently low SNRs (note that, the capacity values in Fig.~\ref{fig:one} are expressed in log scale). The important observation from Fig.~\ref{fig:one} is the improvement in the double-fading channel's capacity at low SNRs with increasing fading severity level (smaller $m$), similar to as alluded to in the low-SNR capacity analysis of a single-hop Nakagami-$m$ channel presented in\!\cite{rezki}.  Particularly at sufficiently low SNRs, we notice a \emph{constant gap/offset} between two capacity curves (when depicted in log-scale) which adjusts as a function of fading severity $m$ - a characteristic quantified in Theorem~\ref{eq:theorem3}. To appreciate the scale of the possible capacity improvement, consider an RF bandwidth of $100$ MHz and $\mathrm{SNR} = -30$ dB; the pinhole channel capacity has almost quadrupled from $500$ Kbps for $m = 4$ (mild) severity to a significant $2$ Mbps for the highest fading severity $m = 0.5$ for the same $\mathrm{SNR}$.

\begin{figure}[htbp]
\vspace*{-1.em}
\centering
\vspace*{-0.2em}
\includegraphics[width=0.5\textwidth]{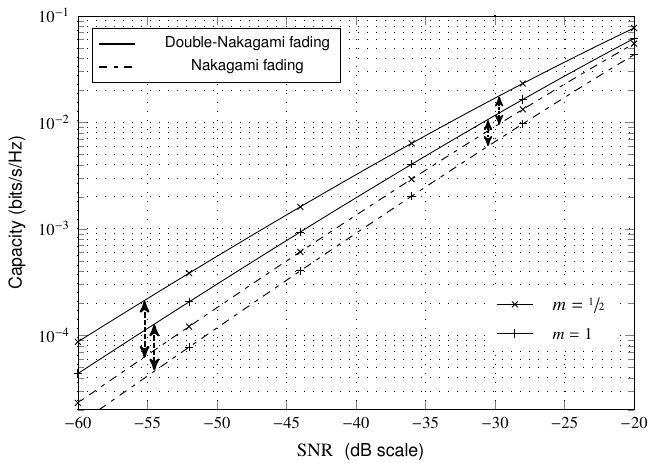}
\vspace*{-1.75em}
\caption{Comparison of exact capacity performance of the Nakagami-$m$ fading (single-hop) and  double-Nakagami-$m$ fading (double-hop) channels at low SNRs. We assume $m_R = m_T = m$ and $\Omega_R = \Omega_T = 1$ for the dyadic channel. Mean channel gain is also normalized to unity for the single-hop channel.}\label{fig:oneB}
\end{figure}

It is also instructive to compare the capacity performance of single-hop and dyadic Nakagami-$m$ fading channels at low SNRs. This is illustrated via numerical results in Fig.~\ref{fig:oneB}: for $m=1$ (Rayleigh case) and $\mathrm{SNR} = -30$ dB, the spectral efficiency has increased from $7\times10^{-3}$ bits/s/Hz to $1.5\times10^{-2}$ bits/s/Hz (almost doubled) when the channel is switched from single-hop to double-hop. For $m = 0.5$ severity level and that the channel is switched from single-hop to dyadic, the 2-fold improvement in capacity still holds at the same $\mathrm{SNR}$. In fact, it can be shown that the capacity improvement factor (comparing cascaded and single-hop fading channels) grows \emph{unbounded} with decreasing $\mathrm{SNR}$. This can be verified to some extent from Fig.~\ref{fig:oneB} at low SNRs: e.g., compare the improvement factor at $\mathrm{SNR} = -30$ dB with that at $\mathrm{SNR} = -55$ dB (for ease of reference, we have placed vertical double-arrow markers at these SNRs in Fig.~\ref{fig:oneB}). In summary, the dyadic Nakagami-$m$ model holds promise to provide for capacity enhancement at low SNR for an increase in fading severity levels.


\begin{figure}[h]
\centering
\includegraphics[width=0.5\textwidth]{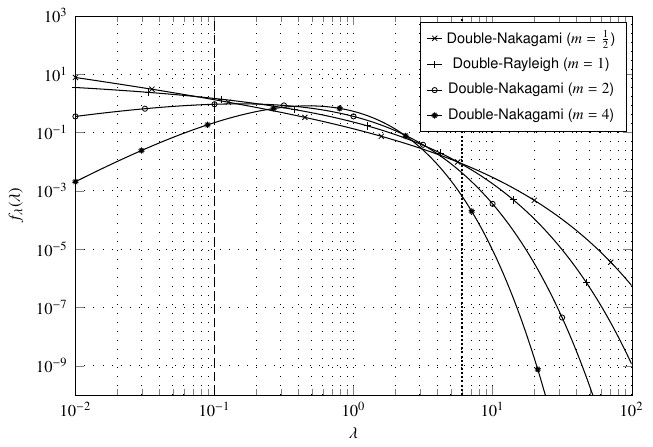}
\vspace*{-1.75em}
\caption{PDF of the double-Nakagami-$m$ pinhole channel gain (see~\eqref{eq:cap:csit0}). For simplicity, we keep $m_R = m_T = m$ (say) and $\Omega_R = \Omega_T = 1$.}\label{fig:oneone}
\end{figure}

To explore the exact reason for the pinhole channel's capacity improvement with increasing fading severity (or decreasing $m$) at low SNRs, we will now argue that it depends crucially on the \emph{tail} of the channel gain distribution $\lambda$. Recall from~\eqref{eq:cutoff1} that decreasing $\mathrm{SNR}$ implies increasing channel cutoff $\lambda_0$ for a given fading statistics; that is, at low SNRs, the transmitter needs to wait until the channel gain reaches its peaks. The larger the probability mass in the tail distribution of the channel gain $\lambda$, the greater the likelihood that the transmitter sees channel peaks in the low-SNR regime. Referring to Fig.~\ref{fig:oneone}, we observe that the probability mass in the tail distribution (e.g., for all values of $\lambda \ge 6$, i.e., right of the thick `dotted' vertical line in Fig.~\ref{fig:oneone}) of the pinhole channel gain $\lambda$ is larger for higher fading severity level (smaller $m$) and vice versa. This tail distribution characteristic has a direct implication: for every fixed $\mathrm{SNR}$ value in the low-SNR regime, the transmitter can choose a larger cutoff $\lambda_0$ for the pinhole channel with higher fading severity to \emph{exploit channel peaks more efficiently} leading to a significant improvement in the spectral efficiency (demonstrated in Fig.~\ref{fig:one}). This heuristic explanation for the channel cutoff adjustment with fading severity in the dyadic pinhole channel model can be readily verified from the arguments in the proof of Theorem~\ref{eq:theorem3}; the channel cutoff $\lambda_0$ (but not the scaled one) decreases inversely proportional to the product $m_T m_R$ (of fading parameters) at low SNRs. Similar arguments with slight modifications can be easily extended to justify the capacity improvement with higher fading severity at low-SNR in a single-hop Nakagami-$m$ fading channel, as illustrated by the numerical results in Fig.~\ref{fig:oneB}.

\noindent
\emph{One last important comment is in order.} Fig.~\ref{fig:oneone} also hints at capacity degradation of the dyadic pinhole channel with fading severity at high SNRs. In the high SNR regime, the channel cutoff $\lambda_0$ is typically low and the impact of transmit power control is minimal\!\cite[Chap.~5]{tsebook}. However, now, the deep fades event, \emph{if statistically significant}, can have an adverse effect on channel capacity. A dyadic pinhole channel typically has a significant probability of deep fades which increases (thus, capacity degrades) further with fading severity. This defining characteristic of the pinhole channel can be inferred directly from Fig.~\ref{fig:oneone} by observing the thickening of the probability mass with decreasing $m$ for low channel gains (e.g., for all values of $\lambda \leq 10^{-1}$; i.e., left of the `dashed' vertical line in Fig.~\ref{fig:oneone}).

\section{Concluding Remarks}\label{sec:conc}
In this work, we presented a deeper characterization of low-rank fading channel's capacity with fading severity in the low-SNR regime. Importantly, we have shown that dyadic (by extension, multi-hop) transmission at low SNRs can leverage higher fading severity for capacity enhancement. To provide more intuition, we argued heuristically that the probability distribution of the channel peaks (tail distribution), in general, plays a crucial role in the capacity performance at very low SNRs. This heuristic argument is then made precise for the dyadic channel by showing that the tail distribution improves with fading severity, which the transmitter operating at low SNR (with full CSI) can exploit to increase spectral efficiency.
\balance

\end{document}